\documentstyle[12pt]{article}
\newcommand{\be}{\begin{equation}}
\newcommand{\ee}{\end{equation}}
\newcommand{\bea}{\begin{eqnarray}}
\newcommand{\eea}{\end{eqnarray}}
\newcommand{\bedm}{\begin{displaymath}}
\newcommand{\eedm}{\end{displaymath}}
\newcommand{\ba}{\begin{array}{c}}
\newcommand{\ea}{\end{array}}
\newcommand{\dis}{\displaystyle}
\newcommand{\ncm}{\newcommand}
\ncm{\ppp}{\P}
\ncm{\qqq}{\S}
\ncm{\eps}{\epsilon}
\ncm{\nn}{\nonumber}
\ncm{\sss}{\scriptstyle}

\input epsf

\begin{document}
\font\twelve=cmbx10 at 15pt
\font\ten=cmbx10 at 12pt
\font\eight=cmr8

\begin{titlepage}

\begin{center}

{\ten Centre de Physique Th\'eorique\footnote{Unit\'e Propre de
Recherche 7061} - CNRS - Luminy, Case 907}

{\ten F-13288 Marseille Cedex 9 - France }

\vspace{1 cm}

{\twelve Gauge Theories with a Layered Phase}

\vspace{0.3 cm}

{\bf A. HULSEBOS\footnote{FB-08, Gau\ss stra\ss e 20, University of
Wuppertal, D-42119, Germany}, C. P. KORTHALS-ALTES\footnote{Centre de
Physique Th\'eorique, CNRS-Luminy, Case 907, F-13288 Marseille cedex
9, France} and S. NICOLIS\footnote{Centre de Physique Th\'eorique,
CNRS-Luminy, Case 907, F-13288 Marseille cedex 9, France {\em and} 
PHYMAT, Universit\'e de Toulon et du Var, F-83957 La Garde cedex, France}}

\vspace{1.5 cm}

{\bf Abstract}

\end{center}

We study abelian gauge theories with anisotropic couplings in $4+D$
dimensions. A layered phase is present, in the absence as well as in
the presence of fermions. A line of second order transitions separates
the layered from the Coulomb phase, if $D\leq 3$.

\vspace{2 cm}


\bigskip

\noindent Number of figures : 9

\bigskip

\noindent May 1994

\noindent CPT-94/P.3036

\noindent hep-th/yymmddd

\bigskip

\noindent anonymous ftp or gopher: cpt.univ-mrs.fr

\begin{tabular}{lc}
\hskip-1truecm World Wide Web: & http://cpt.univ-mrs.fr {\em or}\\
& gopher://cpt.univ-mrs.fr/11/preprints\\
\end{tabular}

\end{titlepage}

\begin{titlepage}
\begin{center}
\begin{Large}

{\bf Gauge Theories with a Layered Phase}

\end{Large}

\vskip 3truecm

A. Hulsebos$^{*}$, C. P. Korthals-Altes$^{**}$ and S. Nicolis$^{**,***}$

\vskip 2truecm 

$^{*}$  {\sl  FB-08, Gau\ss stra\ss e 20, University of Wuppertal, D-42119,
Germany}

\vskip 0.5truecm

$^{**}$ {\sl Centre de Phsyique Th\'eorique, CNRS-Luminy, Case 907\\
F-13288 Marseille cedex 9, France}

\vskip 0.5truecm

$^{***}$ {\sl PHYMAT, Universit\'e de Toulon et du Var, 83957 la Garde
cedex, France}

\vskip 3 truecm

ABSTRACT 

\end{center}

We study abelian gauge theories with anisotropic couplings in $4+D$
dimensions. A layered phase is present, in the absence as well as in
the presence of fermions. A line of second order transitions separates
the layered from the Coulomb phase, if $D\leq 3$.  

\end{titlepage}

\section{Introduction}
The phase structure of gauge theories with isotropic couplings
is known since the advent of lattice gauge theories. For $U(1)$ theories,
there are two phases: a strong coupling and a weak coupling (Coulomb) phase,
above and in four spacetime dimensions, while below
this limit there is only one, strong coupling (confining) phase.
 
It was realized some time ago~\cite{FuNi} that lattice
gauge theories with
anisotropic couplings may have an extra, ``layered'', phase. The
prototype of such a system is the $4+1$--dimensional $U(1)$ theory,  
in which the plaquettes in four dimensions have the same coupling constant
$\beta=1/g^2$,
whereas all plaquettes with links in the fifth dimension have
coupling $\beta'=1/g'^2$.

In the layered phase all four-dimensional hyperplanes, along the fifth
direction, decouple from one
other and in each hyperplane we have a four-dimensional Coulomb
phase. This takes place for $\beta'$ small enough (typically $\leq
1/d$, where $d$ is the dimensionality of the layer, in our case $d=4$)
and for $\beta$ large enough (typically $\beta\geq O(1)$). 

The intuition of Fu and Nielsen~\cite{FuNi} for the occurrence of the
layered phase is  worthwhile bringing up at this point.

Let us suppose that we are in the 5$d$ Coulomb phase (so
$\beta\approx\beta'$ large). Between two test charges we have a
Coulomb force in this phase.  Imagine making
$\beta'$ smaller but keeping $\beta$ fixed. Then the force between the
charges, in the fifth direction, will increase and will become, for
small enough $\beta'$, confining and create a string--whereas the
force will still be Coulomb-like in the four other directions. Thus we
expect the Wilson loops to behave like
\begin{itemize}
\item $W_{\mu\nu}(N,M)\approx \exp-NM\sigma$ (strong coupling);
\item $W_{\nu\mu}(N,M)\approx \exp-(N+M)\tau$ (Coulomb phase);
$1\leq\mu,\,\nu\leq 5$ and
\item 
$W_{\mu\nu}(N,M)=\exp-(N+M)\tau', \,\,\,1\leq\mu,\,\nu\leq 4$;\\
$W_{\mu 5}(N,M)=\exp-NM\sigma'$
(layered phase).
\end{itemize} 
There isn't any layered phase with the roles of $\beta$ and $\beta'$
reversed. 
At small enough $\beta$ one has either strong coupling or
weak coupling and Monte Carlo simulations~\cite{BeRa} seem to
corroborate this picture, though only single plaquette averages were
calculated. This asymmetry is due to the asymmetric role of $\beta$
and $\beta'$: whereas $\beta$ is the coupling restricted to the 4$d$
space, $\beta'$ not only acts in the space orthogonal to the 4$d$
space, but also couples the links in the fifth direction to those
in the others. 

The layered phase is due to the simultaneous presence of Coulomb
forces in the physical direction and confining forces in the other
directions. So we need at least 4 dimensional layers for an Abelian
theory(i.e. $4+D$ spacetime dimensions) and at least 5 dimensions for
a non-Abelian theory ($5+D$ spacetime dimensions).
 
In this paper we will concentrate on two issues: in section 2 on
the order of the various phase transitions and in section 3 on the
presence of fermions, both by mean field techniques and Monte Carlo
methods (section 4). In section 5 we
present results from mean field and Monte Carlo methods and conclusions and outlook are
set forth in section 6. 

\section{Mean Field Theory for 4+D Abelian gauge theory and order of
the transition}

 Mean field techniques for gauge theories~\cite{MF} typically
give  first order transitions--in contrast to spin systems. 
The reason may be found in the structure of the fixed-point equation
for the order parameter, as will be made explicit below and may be
summarized by the remark that a plaquette has four links (gauge
system) while there are two spin variables per link (spin system). 

The standard way to do mean field theory in a $4+D$ anisotropic system
with gauge action 
\begin{eqnarray}
S_{G} & = & \beta\sum_{1\leq\mu<\nu\leq 4, x}
   \left(1-{\Re}U_{\mu\nu}(x)\right) \nn \\
 & + & \beta'{\hskip -0.4truecm}\sum_{\ba {\sss x,\, 1\leq\mu\leq 4,} \\%
{\sss 5\leq\nu\leq 4+D} \ea}
 \left(1-{\Re}U_{\mu\nu}(x)\right) \nn \\
 & + & \beta'{\hskip-0.2truecm}\sum_{5\leq\mu<\nu\leq 4+D,x}
 \left(1-{\Re}U_{\mu\nu}(x)\right)
\label{ga}
\end{eqnarray}
 goes as follows (we use standard notation--the link variables
$U_{\mu}(x)\equiv
e^{{\rm i}\phi_{\mu}(x)}$ in the $x\to x+\mu$ direction, multiplied
around a plaquette give the product $U_{\mu\nu}(x)$). 

The partition function $Z$ is 
\begin{equation}
\label{Z}
Z = \int \prod_{l} U(l)\exp-S_{G}
\end{equation}
and may be trivially rewritten, by inserting the identity operator
\begin{equation}
\label{ident}
1 = \int
\prod_{l}Dv_r(l)Dv_i(l)\;\delta(v_r(l)-{\Re}U(l))\,\delta(v_i(l)-{\Im}U(l))
\end{equation}
and using the integral representation of the delta
function,$\delta(v-U)= \int d\alpha\exp-({\rm i}\alpha(v-U))$. 
The result is 
\begin{eqnarray}
Z & = & \int\prod_{l}DU(l)D\alpha_r(l)D\alpha_{i}(l)Dv_{r}(l)Dv_{i}(l)
\nn \\
 & & \times \; \exp({\rm i}\alpha_r(l){\Re}U(l)+\alpha_i(l){\Im}U(l)) \nn \\
& &\times \; \exp-(S(v)+{\rm i}\alpha_r(l)v_r(l)+{\rm i}\alpha_{i}(l)v_{i}(l))
\label{eff}
\end{eqnarray}
$S(v)$ is obtained by substituting $v_r(l)$ for $\Re U(l)$ and
$v_i(l)$ for $\Im U(l)$ in eq.~(\ref{ga}). 

The $U(l)$ integrations now decouple, at the expense of introducing
the new link variables $\alpha(l)$, and give a contribution 
$\propto\sum_{l}\log\left(I_0\left(\sqrt{\alpha_{r}^{2}(l)+\alpha_{i}^2(l)}\right)\right)$
to the exponential in eq.~(\ref{eff}), leading to an effective action
\begin{equation}
\label{effa}
S_{{\rm eff}} = S(v)+\sum_{l}\log\left(I_0\left(\sqrt{\alpha_{r}^{2}(l)+\alpha_{i}^2(l)}\right)\right)
\end{equation}

Traditionally one introduces axial gauge in the $4$ 
direction\footnote{This is a gauge choice
that allows one to compute the free energy $-\log Z$ easily. Formally,
axial gauge would be $U_{4+D}(x)\equiv 1$ for all points $x$--but the mean
field equations become less transparent.},
$U_{4}(x)\equiv 1$  for all points. 

The equations of motion for the fields $v(l)$ and $\alpha(l)$ may now
be explicitly written down. The correct solution is the configuration
that minimizes the free energy, $-\log Z$. 
Fu and Nielsen~\cite{FuNi} guessed that $\alpha_i(l)=v_i(l)=0$ for all
links $l$ and, furthermore, $v_{\mu}(x)\equiv v$ for
$\mu=1,\ldots,3$, while $v_{\mu}(x)\equiv v'$ for $\mu=5,\ldots,4+D$,
where $v,v'$ are independent of $x$. With this simplifying {\em
Ansatz} the equations of motions become
\begin{equation}
\label{eom1}
\begin{array}{c}
4\beta v^3+2D\beta'v'^2v+2\beta v = \alpha\\
 \\
v = I_1(\alpha)/I_0(\alpha) \\
\end{array}
\end{equation}
\begin{equation}
\label{eom2}
\begin{array}{c}
6\beta'v^2v'+2\beta'(D-1)v'^3+2\beta' v' = \alpha'\\
 \\
v' = I_1(\alpha')/I_0(\alpha')\\
\end{array}
\end{equation}
and the free energy/site (for the $D=1$ case, on which we focus in the 
remainder of this section) may be written as
\begin{equation}
\label{free}
f = 3(-\beta v^2(v^2+1)+\alpha v-\log I_{0}(\alpha))+
\beta'(-3v^2v'^2-v'^2)+\alpha' v' -\log I_{0}(\alpha')
\end{equation}
and delivers plaquette expectation values through 
\begin{equation}
\begin{array}{c}
\dis\frac{\partial}{\partial\beta}f=\sum_{\mu,\nu=1}^{4}\langle
U_{\mu\nu}\rangle\\
\dis\frac{\partial}{\partial\beta'}f=\sum_{\mu=1}^{4}
\langle U_{\mu 5}\rangle\\
\end{array}
\end{equation}
In ref.~\cite{FuNi} these equations were analyzed and three r\'egimes
identified:
\begin{itemize}
\item $v=0,\,\,v'=0$ (strong coupling, I);
\item $v\neq 0, \, \, v'=0$ (layered phase,II);
\item $v\neq 0, \,\,v'\neq 0$ (weak coupling,III).
\end{itemize}
The transition on the axis $\beta=0$ is an artifact of mean field
theory--along this line the theory splits into a sum of 
one-dimensional spin theories. 
Mean field theory is known to break down in low dimensions and
corrections shift the ``transition point'' significantly to the right~\cite{FuNi}. 

Let us now argue the order of the transition(s) from mean field theory.

The transitions  separating I and II, I and III are first order. This
is the expected result from mean field theory, since the $v^3$ term in
eq.~(\ref{eom1}) leads to a jump from $v=0$ (I) to $v\neq 0$ (II,III).
For the strong to layered transition, our result is in agreement with
the latest Monte Carlo data, obtained for $\beta'=0$, i.e.
four-dimensional compact $U(1)$ theory~\cite{MonteCarlo}. 
Along the (II,III) line, however, $v$ stays non-zero; the transition
is marked by $v'$ becoming non-zero. The mean field
equation~(\ref{eom2}) is now relevant--and for $D=1$ $v'$ enters {\em
linearly}, due to the fact that $U_5$ comes in only quadratically in
the 4+1 dimensional anisotropic model! The transition is {\em second
order}--and Monte Carlo results~\cite{BeRa,HuKANi} do indeed confirm
this. For $D>1$ this argument is no longer valid, since $v'$ enters
cubically--however, the {\em coefficient} of this term is not large
enough to drive the transition first order-at least as long as $D\leq
3$. Monte Carlo results are not yet available for the higher
dimensional case ($D>1$), so we cannot check the validity of this mean 
field prediction.

\section{Coupling fermions}
The presence of fermions renders the order parameter inefficient--it
will always indicate perimeter behavior. What one should look at
instead is the fermion spectrum, which is different in the three
phases:
\begin{itemize}
\item bound states with strong coupling (I);
\item fermions move freely in the four-dimensional subspace as in a
four-dimensional Coulomb phase, but feel a confining force in the
extra directions (II);
\item free fermions, positronium states in $4+D$ dimensions. 
\end{itemize}
We have checked, by mean field methods, that the phase diagram
described above is not qualitatively altered by the presence of the
fermions. In this section we will explain these methods.

Our starting point is the fermionic action, with a Wilson term
\begin{eqnarray}
S_{F} & = & -\frac{1}{2} \, \sum_{x,\mu}\overline{\Psi}(x)\gamma_{\mu}
[U_{\mu}(x)\Psi(x+\hat\mu)-U_{\mu}^{\dag}(x-\hat\mu)\Psi(x-\hat\mu)]
\nn \\
& + & \frac{r}{2}\sum_{x,\mu}\overline{\Psi}\left(U_{\mu}(x)\Psi(x+\hat\mu)-
2\Psi(x)+U_{\mu}^{\dag}(x-\hat\mu)\Psi(x-\hat\mu)\right) \nn \\
& + & M \sum_{x}\overline{\Psi}(x)\Psi(x) \nn \\
& \equiv & \sum_{x,x'}\overline{\Psi}(x){\cal M}(x,x')\Psi(x')
\label{fa}
\end{eqnarray}
The total action is the sum of the above and the gauge
action, equation (\ref{ga})
\begin{equation}
\label{tota}
S = S_{G}+S_{F}
\end{equation}
The partition function now reads
\begin{equation}
\label{fZ}
Z = \int \prod_{x}D\overline{\Psi}(x)D\Psi(x)\prod_{l}DU(l)\exp-S
\end{equation}
The fermions may then be integrated out, leading to the effective
action
\begin{equation}
\label{effaf}
S_{{\rm eff}}\equiv S-{\rm Tr}\log{\cal M}(U)
\end{equation}
The mean field approximation goes through exactly the same steps as in
the pure gauge case, leading to
\begin{equation}
\label{fZZ}
Z=\int_{{\rm -i}\infty}^{{\rm i}\infty}\prod_{l}D\alpha(l)
\int Dv(l)\exp-\left(S_{{\rm eff}}+\alpha v-\log
I_{0}(\alpha)\right)
\end{equation}
The difference with the pure gauge case is that the fermionic
determinant leads to a non-local action for the $v$-variables. The
equations of motions now become
\begin{equation}
\label{eom1f}
\begin{array}{c}
4\beta v^3+2D\beta'v'^2v+2\beta v + j(v,v')= \alpha\\
 \\
v = I_1(\alpha)/I_0(\alpha) \\
\end{array}
\end{equation}
and
\begin{equation}
\label{eom2f}
\begin{array}{c}
6\beta'v^2v'+2\beta'(D-1)v'^3+2\beta' v' + j'(v,v')= \alpha'\\
 \\
v' = I_1(\alpha')/I_0(\alpha')\\
\end{array}
\end{equation}
 The currents $j(v,v')$ and $j'(v,v')$ are short-hand for 
\begin{equation}
\label{current}
j(v,v')=\frac{\delta}{\delta v_{\mu}(x)}{\rm Tr}
\log{\cal M}|_{v_{\mu}(x)=v,v_{\nu}(x)=v'},
 \,\,\,\mu=1,2,3;\,\,\nu=5,\ldots,4+D
\end{equation}
and
\begin{equation}
\label{current5}
j'(v,v')=\frac{\delta}{\delta v_{\nu}(x)}{\rm Tr}
\log{\cal M}|_{v_{\mu}(x)=v,v_{\nu}(x)=v'},
 \,\,\,\mu=1,2,3;\,\,\nu=5,\ldots,4+D
\end{equation}
So $j(v,v'),\,\,j'(v,v')$ are the expectation values of the components
of the fermionic current in the
background field $(v,v')$. 
Straightforward manipulations lead to the expressions 
\begin{eqnarray}
\label{currents}
j_{\mu}(v,v') & = & {\rm Tr}\left(\frac{\gamma_{\mu}+r}{2}
  G_{x,x+\hat\mu}(v,v')\right) \nn \\
 & + & {\rm Tr}\left(\frac{-\gamma_{\mu}+r}{2}
  G_{x+\hat\mu,x}(v,v')\right), \,\,\mu=1,2,3,5,\ldots,4+D
\end{eqnarray}
where $G(v,v')$ is the fermion propagator in the background $(v,v')$
and the trace is only over Dirac indices.

Since the background is constant in space, the currents $j,j'$ can be
computed implicitly as functions of $v,v'$ through a $4+D$ dimensional
momentum integral.

\begin{figure}[tbp]
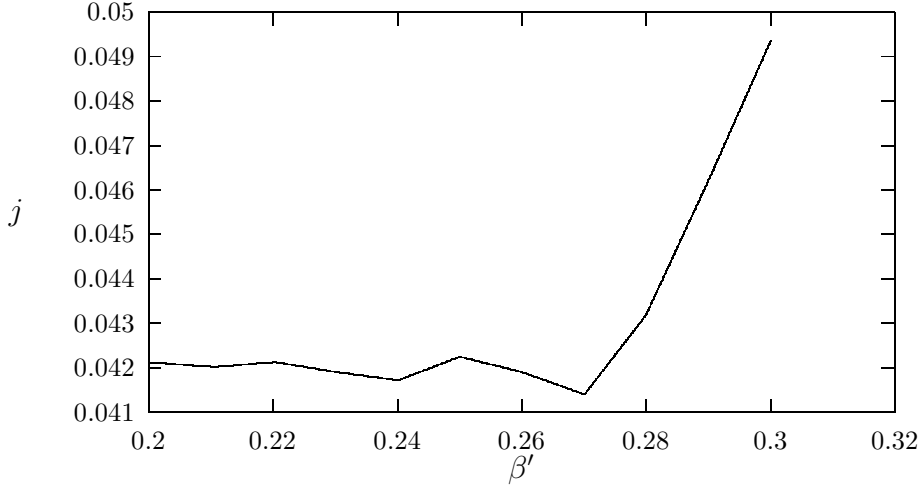

\input J.tex
\caption[]{The current $j$, as a function of $\beta'$, for
$\beta=1.2$.}
\label{J}
\end{figure}

\begin{figure}[tbp]
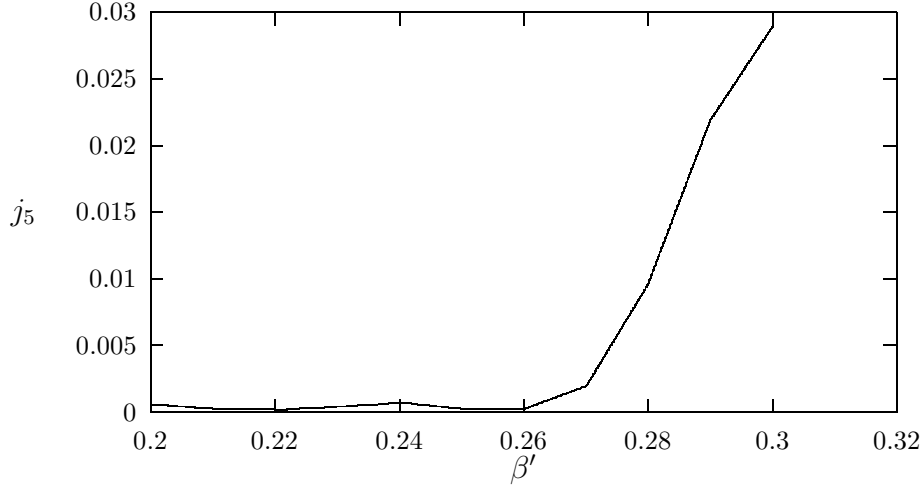

\input J5.tex
\caption[]{The current $j_5$, as a function of $\beta'$, for
$\beta=1.2$.}
\label{J5}
\end{figure}

Let us now introduce the variables
\begin{equation}
\label{wilson}
W\equiv M-r\left(\sum_{\lambda=1}^{3}(1-v\cos p_{\lambda})+1-\cos p_{4}+
\sum_{\lambda=5}^{4+D}(1-v'\cos p_{\lambda})\right)
\end{equation}
and
\begin{equation}
\label{denom}
P\equiv \sum_{\lambda=1}^{3}v^2\sin^2 p_{\lambda}+\sin^2 p_4 +
\sum_{\lambda=5}^{4+D}v'^2\sin^2 p_{\lambda} + W^2
\end{equation}
in terms of which the currents may be written as
\begin{equation}
\label{currentI}
j_{\mu}=4\int_{-\pi}^{\pi}\frac{d^{4+D}p}{(2\pi)^{4+D}}\left[
v\sin^2 p_{\mu} + r\cos p_{\mu} W\right]\frac{1}{P}
\hspace*{1cm} \mu=1,2,3
\end{equation}
and
\begin{equation}
\label{current5I}
j_{\nu}=4\int_{-\pi}^{\pi}\frac{d^{4+D}p}{(2\pi)^{4+D}}\left[
v'\sin^2 p_{\nu} + r\cos p_{\nu} W\right]\frac{1}{P}
\hspace*{1cm} \nu=5,\ldots,4+D
\end{equation}
Obviously, if $v'=0$ then $j'=0$ (c.f.\ equation (\ref{eom2f})--or,
more explicitly, by combining (\ref{wilson}) and (\ref{current5I}))
 and the equations of motion are
identical with or without fermions. Assuming a second order transition
between the layered and the Coulomb phases we find 
the shift of the transition point induced by the presence of the
fermions (for $v=1$) as
\begin{equation}
\label{shift}
\beta'_{c}=\frac{1}{4}-\frac{\delta}{8}
\end{equation}
where
\begin{equation}
\label{shiftI}
\delta=\frac{1}{\pi}\int_{\pi}^{\pi}\frac{d^4p}{(2\pi)^4}\left[\frac{(1+r^2)S^2+(1-r^2)W^2}{P^2}\right]
\end{equation}
where $S=\sum_{\mu=1}^{4}\sin^2 p_{\mu}$, $W$ and $P$ as above, but
for $v'=0,v=1$. Note that the shift is {\em negative} and very small (except for special values of $M$ where
$P$ has poles), i.e. the
layered phase, in the presence of fermions, loses ground against the
Coulomb phase. 
 Physically the sign may be understood from the fact
that the presence of fermions weakens the confining forces,
responsible for the layered phase\footnote{In our original
paper~\cite{KANP} there is an unfortunate error in the drawing of the
phase diagram.}. Thus there is a critical number of fermions, beyond
which there isn't any layered phase. 
The fermionic currents $j$ are independent of $\beta'$ in the layered
phase (since $v$ is). The current $j'$ is identically zero in that
phase. For fixed $\beta$ it grows in the Coulomb phase, cf.
figs.~\ref{J},~\ref{J5}.

\section{Monte Carlo Methods}
The size of the five dimensional system was taken to be $4^5$ and
$8^5$. 
For the pure gauge simulations we used a 5-hit Metropolis algorithm with 
dynamically adjustable stepsizes $\eps_{\mbox{\tiny sp}}$ and $\eps_5$ in 
order to maintain an acceptence rate of roughly 50\% in each direction.

As we are interested in the phase structure, we performed three thermal runs on
each lattice size. Such a run consisted of 5000 updates at the starting 
values of the couplings $\beta$ and $\beta'$ for thermalization, 
followed by 1000 updates and measurements,
after which either $\beta$ or $\beta'$ was increased
by 0.01, and another 1000 updates and measurements were made without 
rethermalization. After having reached the final values of $\beta$ and $\beta'$,
the process is reversed intil the starting values have been reached again.
There were no thermalization sweeps done at the start of the second leg of
the run.

The number of 1000 measurements was chosen with the following reasoning in 
mind. In case of a first order phase transition, there are, loosely speaking,
two autocorrelation times at play: the first being the tunnelling time 
$\tau^{\mbox{\tiny T}}$, and the second being the relaxation times
$\tau^{\mbox{\tiny  S}}_i$ within the states $i$. The number of measurements,
$N_{\mbox{\tiny  meas}}$, should then satisfy
\bedm
\tau^{\mbox{\tiny  S}}_i < N_{\mbox{\tiny  meas}} \ll \tau^{\mbox{\tiny  T}}.
\eedm
For second and higher order phase transitions, $\tau^{\mbox{\tiny  T}}$ is
absent, and we are left with the condition
\bedm
\tau^{\mbox{\tiny  S}}_i < N_{\mbox{\tiny  meas}}.
\eedm

Although changing the couplings will bring the system out of equilibrium, we 
might hope that this disturbance is relatively small, and the system will
relax to equilibrium within $\tau^{\mbox{\tiny  S}}$ sweeps.

As observables to measure we choose the average plaquettes
$U_{\mbox{\tiny  sp}}$ and $U_5$,
\bea
U_{\mbox{\tiny  sp}} & = & \Biggl< \frac{1}{6V} 
\sum_{\ba{\sss x,\mu,\nu}\\{\sss \mu<\nu<5}\\ \ea}
 \Re U_{\mu\nu}(x) \Biggr>  \\
U_5 & = & \Biggl< \frac{1}{4V} \sum_{\ba{\sss x,\mu}\\{\sss \mu<5}\\
 \ea} \Re U_{\mu 5}(x) \Biggr> 
\eea
and the Polyakov line correlators $P_{\mbox{\tiny  sp}}$ and $P_5$
\bea
P_{\mbox{\tiny  sp}} & = & \Biggl< \frac{1}{6V} \sum_{ \ba{\sss x,\mu}\\{\sss
 \mu<\nu<5}\\ \ea} \Re \Bigl( p^{\dag}_{\mu}(x)
p_{\mu}(x+{\textstyle \frac{N}{2}} \hat{\nu})\Bigr) \Biggr>  \\
P_5 & = & \Biggl< \frac{1}{4V} \sum_{\sss x,\mu}\Re\Bigl( p^{\dag}_\mu(x)
p_5(x+{\textstyle \frac{N}{2}} \hat{\mu}) \Bigr) \Biggr> 
\eea
where $p_{\mu}(x)$ is the Polyakov line starting at point $x$, running in the
$\hat{\mu}$ direction and $N$ is the lattice size in the corresponding
direction. 
As the phase transitions in this model are conjectured
to be confining-deconfining transitions in one or more dimensions, these
observables are indeed order parameters for these phase transitions. For a
confining phase, $P$ will be zero as the system will be disordered over
long distances. In a Coulombic phase, however, $P$ will be non-zero, as now 
such long correlations are allowed.

When including fermions, we used a  Hybrid Monte Carlo algorithm with a first
order discretization of Hamilton's equations. A standard Conjugate Gradient
algorithm was used for matrix inversion. One sweep consisted of
momentum refreshment, followed by 10 integrations with stepsize $\tau = 0.05$.
An accept/reject step was done in order to maintain balance. The runs were
performed as in the pure gauge case, but now with 1500 thermalization sweeps
and 250 sweeps at each $(\beta,\beta')$ pair.

The actual simulations were done on the Wuppertal CM-5 in its various 
incarnations. We are, therefore, unable to make comments about the performance
of the two programs on this machine.

\section{Results}
This section will be subdivided into two parts: one for the pure gauge
case and one including the fermions. For each case we shall confront
results (namely plaquette values) from mean field theory with Monte
Carlo simulations-so we work in five spacetime dimensions in what
follows and will comment on what happens for $D=2,3$ at the end.

\subsection{Pure gauge results}
In this paragraph we report on mean field calculations and Monte Carlo
simulations for the pure gauge case. The quantities of interest are
the average plaquettes.

\begin{figure}
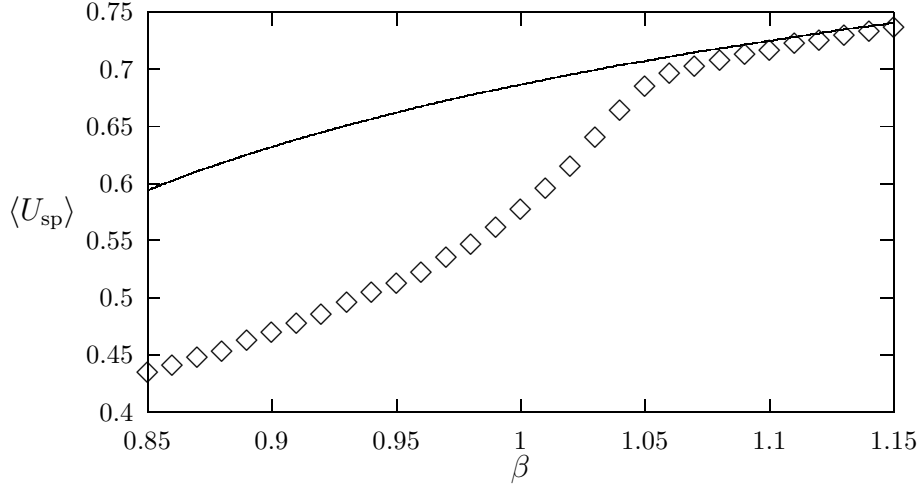

\input s2lg_paper.tex
\caption[]{Average space-space plaquette {\em vs.} $\beta$ at
$\beta'=0.2$. Solid line is mean field theory, points are Monte Carlo
results from 1000 measurements on an $8^5$ lattice. }
\label{plaquette}
\end{figure}

Using the equations of motion, (\ref{eom1}) and (\ref{eom2}), one obtains the following expressions,
within mean field theory
\begin{equation}
\label{plaq}
\begin{array}{c}
\langle U_{\rm sp}\rangle=-\dis\frac{1}{6}\frac{\partial}{\partial\beta}f=\frac{v^4+v^2}{2}\\
\langle U_{\mu 5}\rangle=-\dis\frac{1}{4}\frac{\partial}{\partial\beta'}f=\frac{v'^2(3v^2+1)}{4}\\
\end{array}
\end{equation}
These reflect the fact that the equations of motion cancel all {\em
implicit} dependence of the free energy on $\beta$ and $\beta'$. So
the rhs of eq.~(\ref{plaq}) represents only the {\em explicit}
dependence and is, therefore, equally valid when fermions are included.
In
fig.~\ref{plaquette} we display typical plaquette values {\em vs.}
$\beta$ at fixed $\beta'$, relevant for the
strong\ $\leftrightarrow$ layered transition. Mean field theory predicts a 
transition at $\beta=0.8$, while the simulations indicate that the transition is at
$\beta=1.02$. It should be noted that corrections to mean field theory
also indicate an upward shift in the transition point~\cite{FuNi}. 
In the layered phase, the mean field prediction is that the
space-space plaquette, for fixed $\beta$, doesn't depend on $\beta'$. 
The reason is that in the
layered phase $v'=0$ and $\beta'$  enters the mean field equations
only in conjunction with it; furthermore, from eq.~(\ref{plaq}), the plaquette in the fifth
direction is zero.
There is, however, a strong coupling correction $\delta U =
\beta'/2+O(\beta'^2)$ that must be added-this is a well-known result
that may also serve as a check on the simulations.

\begin{figure}
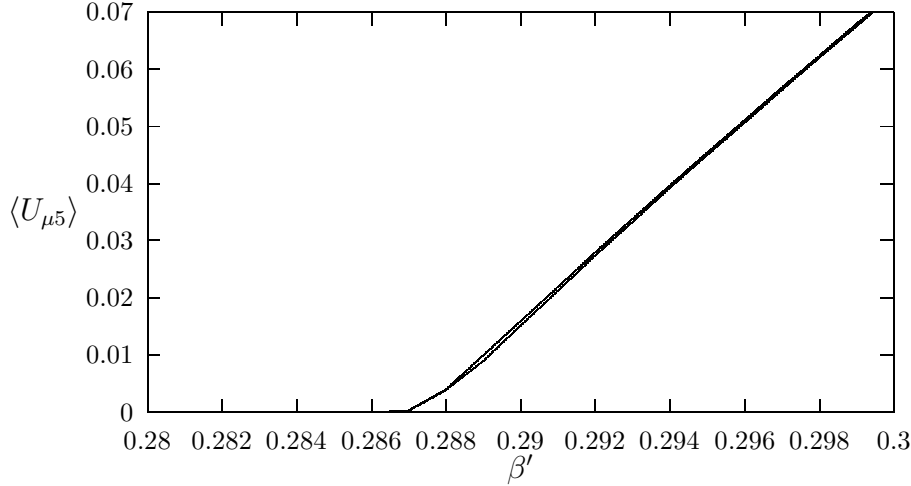

\input l2w1.2hilot5.tex
\caption[]{Hysteresis loop for the layered\ $\leftrightarrow$ Coulomb
transition at $\beta=1.2$. Mean field theory.}
\label{hysteresis}
\end{figure}

\begin{figure}
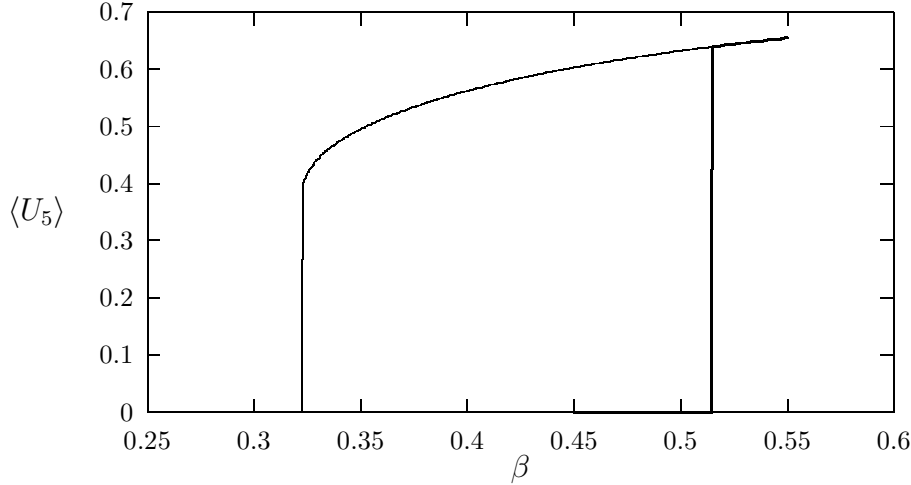

\input s2whyst1
\caption[]{Hysteresis loop for the strong\ $\leftrightarrow$ Coulomb
phase transition at $\beta'=1$.}
\label{hysteresis1}
\end{figure}

Finally, we present evidence that the
layered\/$\leftrightarrow$\/Coulomb is second order, namely hysteresis
loops, in fig.~\ref{hysteresis}. As a comparison, we display
hysteresis loops for the strong\ $\leftrightarrow$ Coulomb transition
in fig.~\ref{hysteresis1}.  

\begin{figure}
\epsfxsize=8truecm
\epsfbox{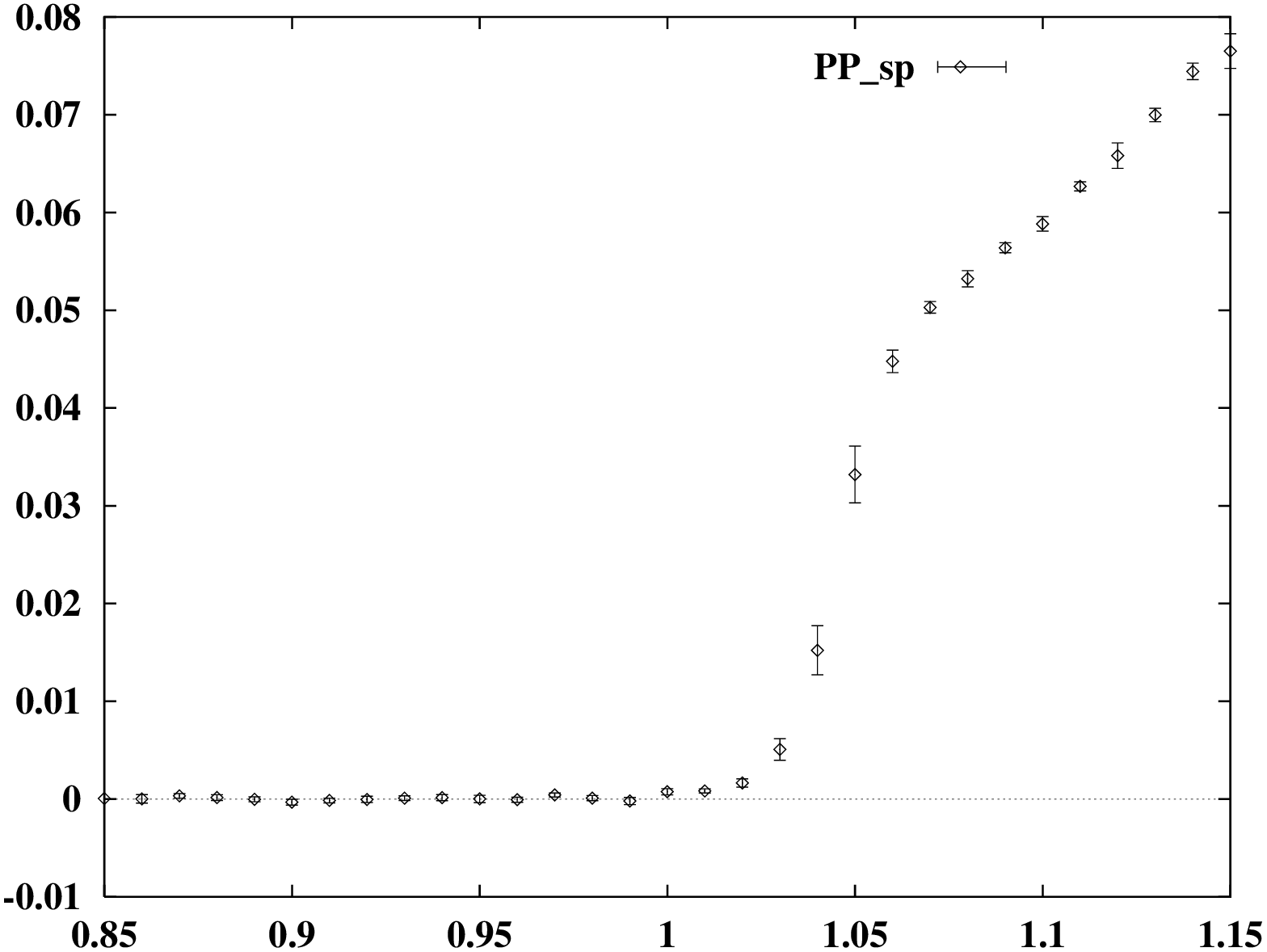}
\caption[]{Spatial Polyakov line correlators  at $\beta'=0.2$ from an $8^5$ lattice}
\label{Polyakov}
\end{figure}

For the strong\/$\leftrightarrow$\/layered transition we also display 
spatial Polyakov line data in Figs.~\ref{Polyakov}.
These data are the first Monte Carlo data using the order parameter,
not just plaquette expectation values, that show the occurrence of the
layered phase. 
Our results may be summarized by the phase diagram for the pure gauge
theory, displayed in fig.~\ref{phadi5}. The main novelty is the
evidence that the layered\ $\leftrightarrow$ Coulomb transition is
second order, in agreement with the intuitive argument of the
previous section. For higher dimensions of spacetime, relevant for
other theories, the situation is the following: mean field theory
predicts that the layered\ $\leftrightarrow$  Coulomb phase transition
stays second order for $4+D\leq 7$, while it becomes first order for $4+D>7$;
this is based on a calculation of
the hysteresis loops. Furthermore, as may be seen from
fig.~\ref{phadi6}, the strong coupling phase, for $4+D=6$ is
significantly reduced, in favor of the Coulomb phase. Unfortunately,
Monte Carlo data are not available to check whether this is an
artifact of the mean field approximation. 

\begin{figure}
\input phadi5
\caption[]{Phase Diagram of the five dimensional, $U(1)$ pure gauge
theory.}
\label{phadi5}
\end{figure}    

\begin{figure}
\input phadi6
\caption[]{Phase Diagram of the six dimensional, $U(1)$ pure gauge
theory.}
\label{phadi6}
\end{figure}

Of course there are several caveats that must be kept in mind:
(a) corrections to mean field theory may be important--the analysis of
ref.~\cite{FuNi} indicates that they do not lead to qualitative
changes in five dimensions, but are significant in three dimensions;
regarding these corrections, it should be noted that they tend to {\em
soften} the first order transitions usually encountered~\cite{Zuber};
(b) regarding the numerical simulations, the analysis of
ref.~\cite{MonteCarlo}of the order of the transition is certainly relevant here--the results
presented are preliminary, certainly indicative of an issue that
requires more extensive work~\cite{HuKANi} and, of course, a very long
correlation length cannot be ruled out.

\subsection{Fermions}
We use the Wilson action, as discussed in the previous section, with
one fermion flavor, and
set $r=1$ and $M=1$. The entire analysis goes through intact-the only
calculational complication arising from the fact that every iteration
of the mean field equations towards the fixed point requires the
numerical evaluation of a four (in the layered phase) or five
dimensional integral for which  we used the Vegas routine~\cite{LePage}.
Computing, once more, the hysteresis loops
through the transitions, we find that the presence of the fermions
doesn't change the order of the three transitions. Consistent with
this result we find a shift of the layered\ $\leftrightarrow$  Coulomb
transition in favor of the Coulomb phase.

\begin{figure}
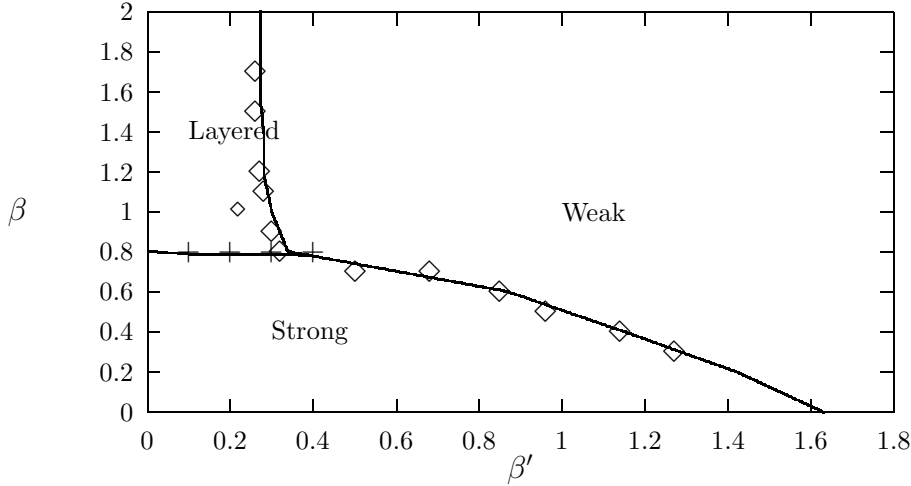

\input fermions3.tex
\caption[]{Phase diagram including fermions of mass $M=1$ (points).
Mean field theory predicts the strong to layered transition at
$\beta=0.8$; Monte Carlo estimates it at 1.02. 
The pure gauge phase diagram is included for comparison (solid
line). }
\label{fermions3}
\end{figure}

To check these predictions we have performed Monte Carlo simulations,
measuring the same observables as in the pure gauge case,
using the Hybrid Monte Carlo method and our results are summarized in
the phase diagram in fig.~\ref{fermions3}.
The caveats stated above apply here as well. High statistics
simulations are needed to clarify whether the results obtained for the
layered\ $\leftrightarrow$  Coulomb transition do indeed hold up. The
claim is that the problem deserves this effort~\cite{HuKANi}.

\subsection{Continuum Limit}

The presence of the line of {\em second order} transitions in the
phase diagrams 
is intriguing. On both sides of this line we have a
phase with zero mass particles and a Coulomb law; the difference is
that the force falls off like $1/r^2$ 
in the four physical directions in
the layered phase and like $1/r^{2+D}$ in all $4+D$ directions in the weak
coupling phase.

In the layered phase the Wilson loops in the extra directions obey an
area law. At $\beta'=0$ the string tension becomes infinite and no
correlation in these directions survives--we have a stack of decoupled $4d$
layers, each of unit thickness (for $D=1$ or ``volumes'' for $D=2,3$).
At $\beta'>0$ , but $\beta'<\beta'_c(\beta)$ (so $v'=0$), one has
finite, but exponentially damped, correlations between the layers. 

Presumably a continuum limit of this system exists at the tricritical
point in figs.~\ref{phadi5},\ref{phadi6},\ref{fermions3}. At $\beta'=0$ we cannot
take any continuum limit~\cite{MonteCarlo}.

\section{Conclusions and outlook}
We have presented evidence, analytical (within the mean field
approximation) and numerical (through Monte Carlo simulations) that
gauge theories, with anisotropic couplings, naturally support layered phases
that admit a non trivial continuum limit. 

Our results show unambiguously a line of second order phase
transitions, between layered and Coulomb phase in fig.~\ref{phadi5}. 
Furthermore, this statement
remains true in the presence of fermions, fig.~\ref{fermions3}. The implications of these findings are quite broad in
scope. 
In the same context one may find
a natural setting for
investigations of membranes and models with chiral fermions. For this
last case the present calculations are a necessary first step for the
location of the chiral layer~\cite{KANP,NarNeub}.  

In this paper we only looked at abelian theories, in $4+D$ dimensions.
If we heat up the system, one can include non-abelian systems, as
soon as they drop into the deconfined phase. We would expect our
results for the Polyakov loops to remain qualitatively the same--with
the caveat that {\em only} the Polyakov loop in the Euclidean time
direction deconfines.
Systems of this kind have been studied by other means in
ref.~\cite{LiangYingTian}.

\section*{Acknowledgements}
We would like to thank Wolfgang Bock, Michael Creutz, Ivan Horv\'ath,
Holger B.~Nielsen, Claudio Parrinello and Jan Smit for useful 
discussions. 
Wolfgang Bock is also thanked for making his Hybrid
Monte Carlo code available to us.
The work of AH was supported by EC contract SC1 *CT91-0642.

\end{document}